\documentstyle[12pt,epsfig]{article}
\def\lbldef#1#2{\expandafter\gdef\csname #1\endcsname {#2}}

\def\href#1#2{#2}  
\begin{document}
\baselineskip=15.5pt
\pagestyle{plain}
\setcounter{page}{1}
%\renewcommand{\thefootnote}{\fnsymbol{footnote}}
%--------+---------+---------+---------+---------+---------+---------+
%Title page

\begin{titlepage}

\begin{flushright}
IC/2000/167\\
hep-th/0011024
\end{flushright}
\vspace{10 mm}

\begin{center}
{\Large Brane Cosmology in the Background of D-Brane \\
with NS $B$ Field}

\vspace{5mm}

\end{center}

\vspace{5 mm}

\begin{center}
{\large Donam Youm\footnote{E-mail: youmd@ictp.trieste.it}}

\vspace{3mm}

ICTP, Strada Costiera 11, 34014 Trieste, Italy

\end{center}

\vspace{1cm}

\begin{center}
{\large Abstract}
\end{center}

\noindent

We study the cosmological evolution of the four-dimensional universe 
on the probe D3-brane in geodesic motion in the curved background of 
the source D$p$-brane with non-zero NS $B$ field.  The Friedman 
equations describing the expansion of the brane universe are obtained 
and analyzed for various limits.  We elaborate on corrections to the 
cosmological evolution due to nonzero NS $B$ field.

\vspace{1cm}
\begin{flushleft}
November, 2000
\end{flushleft}
\end{titlepage}
\newpage

\section{Introduction}

Recently, theorists have actively studied the possibility that our 
four-dimensional universe may be a three-brane embedded in higher-dimensional 
spacetime, motivated by the recent proposals on solving the hierarchy 
problem with the large extra dimensions \cite{add1,aadd,add2} and 
through the Randall-Sundrum (RS) scenario \cite{rs1,rs2,rs3}.  In 
such brane world scenario, all the matter fields live on a brane, while 
gravity can propagate in the bulk space.  This idea is also reminiscence 
of the earlier proposed Horava-Witten picture \cite{hw} for the 
$E_8\times E_8$ heterotic string theory, where matter is regarded as 
being confined on two ten-dimensional hypersurfaces on the boundaries 
of the eleven-dimensional spacetime.  

A lot of work on the cosmological models based on the brane universe 
idea has been done recently.  Most of the work takes an approach that 
the cosmological evolution of the universe is due to the time evolution 
of energy density on the brane, e.g. \cite{dt,bdl,cgs,cgkt,cf,ftw,kkop}.  
(See also Ref. \cite{ano} for another approach of cosmology the RS 
scenario, where the canonical Wheeler-DeWitt formalism is adopted.) 
In this paper, we follow a different approach based on the idea that the 
cosmological evolution of our four-dimensional universe is due to the 
geodesic motion of the (probe) universe 3-brane in the curved background 
of other branes in the bulk \cite{cr,cpr,kr,kk}.  In this approach, the 
motion in ambient space induces cosmological expansion or contraction 
of the four-dimensional universe on the probe 3-brane, simulating various 
kinds of matter or a cosmological constant responsible for cosmological 
expansion \cite{kk}.  Thereby, the cosmological models based on such approach 
are dubbed as mirage, meaning that the cosmological expansion is not due to 
real matter or energy density on our universe brane but due to something 
else.  Following this approach, the cosmological evolution of the brane 
universe in various string theory backgrounds was studied 
\cite{pp1,kim1,kim2,pp2}.  

In this paper, we study the cosmological evolution of the brane universe on 
the probe D3-brane moving in the background of source D-brane with nonzero 
NS $B$ field.  The D-brane background with nonzero NS $B$ field is 
particularly interesting because of its relevance to the recently revived 
noncommutative theories \cite{cds,dh,sw}.   The mirage cosmology in such 
background may thereby serve as a toy cosmological model for providing with 
insight into more realistic cosmological models in noncommutative spacetime.  
We find that noncommutativity generally leads to anisotropicity of 
spacetime.  This is not unexpected, because according to the noncommutativity 
prescription of A. Connes one has to pick two spatial directions or 
one spatial direction along with the time direction associated with the 
noncommutative parameter, thereby loosing the meaning of isotropy 
of three-dimensional space.  We also find that nonzero NS $B$ field, or 
equivalently the noncommutative parameter of the brane worldvolume theories, 
gives rise to nontrivial corrections to the evolution of brane universe 
when the NS $B$ field is electric, whereas the magnetic NS $B$ field gives 
rise to qualitatively same cosmological evolution as the commutative case 
(this is a consequence of the Morita equivalence of noncommuative theories, 
as the cosmological expansion in the mirage cosmology is determined by the 
dynamics of the probe D-brane in the background of the source D-brane).  
It is a subject of future research to understand the physical meaning of 
this result and to study more realistic cosmological models in noncommutative 
spacetime.

The paper is organized as follows.  In section 2, we review the 
general formalism of mirage cosmology proposed in Ref. \cite{kk}.  
In section 3, we study the cosmological evolution on the probe 
D3-brane moving in the background of the source D-brane with nonzero 
magnetic NS $B$ field background.  In section 4, we repeat the same 
analysis with the source D-brane with nonzero electric NS $B$ field.

\section{General Formalism}

In the formalism of mirage cosmology proposed in Ref. \cite{kk}, 
our four-dimensional universe is regarded as the probe D3-brane 
moving freely in the background of some source brane.  
The source brane is assumed to be not affected by the back-reaction 
from the probe D3-brane, as the source brane is much heavier than 
the probe brane.  The metric of the source $p$-brane is parameterized as
\begin{equation}
ds^2_{10}=G_{\mu\nu}dx^{\mu}dx^{\nu}
=g_{00}(r)dt^2+g(r)ds^2_p+g_{rr}(r)dr^2+g_S(r)d\Omega_{8-p},
\label{sourcemet}
\end{equation}
where $\mu,\nu=0,1,...,9$, the line element of the unit sphere is 
parameterized as $d\Omega_{8-p}=h_{ij}(\varphi)d\varphi^id\varphi^j$ 
($i,j=1,...,8-p$) and $p\geq 3$.  We consider the possibility that 
the D-brane configuration contains the NS 2-form field $B_{\mu\nu}$ and 
the D-brane worldvolume $U(1)$ gauge field $A_{\mu}$ with the field 
strength $F_{\mu\nu}$, which cause noncommutativity on the D-brane 
worldvolume.  The action for the probe D3-brane is given by
\footnote{I would like to thank R. Cai for pointing out the last term 
in the probe D3-brane action that was missing in the first version of 
the paper.}
\begin{eqnarray}
S&=&T_3\int d^4\xi\,{\cal L}=T_3\int d^4\xi\,e^{-\phi}\sqrt{-{\rm det}
\left(\hat{G}_{\alpha\beta}+2\pi\alpha^{\prime}F_{\alpha\beta}-
\hat{B}_{\alpha\beta}\right)}
\cr
& &+T_3\int d^4\xi\,\hat{A}^4+T_3\int d^4\xi\,\hat{A}^2\wedge \hat{B},
\label{probact}
\end{eqnarray}
where the hatted fields are pullbacks of the target space fields, defined 
for example for the case of the metric as
\begin{equation}
\hat{G}_{\alpha\beta}=G_{\mu\nu}{{\partial x^{\mu}}\over{\partial\xi^{\alpha}}}
{{\partial x^{\nu}}\over{\partial\xi^{\beta}}},
\label{hatflds}
\end{equation}
where $\alpha,\beta=0,1,...,p$.  
We assume the static gauge $x^{\alpha}=\xi^{\alpha}$ for the probe action 
and that the transverse target space coordinates depend on the time 
coordinate $\xi^0=t$, only.  Note, the fields in the probe action 
(\ref{probact}) correspond to the fields generated by the source brane, 
as felt by the probe D3-brane.  
Substituting the solution for the source brane into the probe action, 
generally one has the following form of the probe Lagrangian density
\begin{equation}
{\cal L}=\sqrt{A(r)-B(r)\dot{r}^2-D(r)h_{ij}\dot{\varphi}^i\dot{\varphi}^j}
-C(r),
\label{lagden}
\end{equation}
where the overdot stands for derivative with respect to $t$ and the 
expressions for $A$, $B$, $C$ and $D$ depend on the type of the source 
brane.  

To obtain the equations describing the geodesic motion of the probe brane, 
we consider the following canonical momenta and the Hamiltonian of the 
probe brane:
\begin{eqnarray}
p_r&=&{{\partial{\cal L}}\over{\partial\dot{r}}}=
-{{B\dot{r}}\over\sqrt{A-B\dot{r}^2-Dh_{ij}\dot{\varphi}^i\dot{\varphi}^j}},
\cr
p_i&=&{{\partial{\cal L}}\over{\partial\dot{\varphi^i}}}=
-{{Dh_{ij}\dot{\varphi}^j}\over\sqrt{A-B\dot{r}^2-Dh_{ij}\dot{\varphi}^i
\dot{\varphi}^j}},
\cr
H&=&-E=p_r\dot{r}+p_i\dot{\varphi}^i-{\cal L}=C-{A\over\sqrt{A-B\dot{r}^2
-Dh_{ij}\dot{\varphi}^i\dot{\varphi}^j}}.
\label{momham}
\end{eqnarray}
Making use of the fact that the energy $E$ and the total angular momentum 
$h^{ij}p_ip_j=\ell^2$ are conserved, from Eq. (\ref{momham}) one obtains 
the following equation describing the radial motion of the probe:
\begin{equation}
\dot{r}^2={A\over B}\left(1-{A\over{(C+E)^2}}{{D+\ell^2}\over D}\right),
\label{radmtn}
\end{equation}
along with
\begin{equation}
h_{ij}\dot{\varphi}^i\dot{\varphi}^j={{A^2\ell^2}\over{D^2(C+E)^2}}.
\label{angmtn}
\end{equation}
From Eq. (\ref{radmtn}), one obtains the following constraint:
\begin{equation}
1-{A\over{(C+E)^2}}{{D+\ell^2}\over D}\geq 0,
\label{cnstrnt}
\end{equation}
which restricts the allowed value of the scale factor $a$ defined below. 

The metric of the four-dimensional universe in the mirage cosmology 
scenario is given by the following induced metric on the probe D3-brane:
\begin{equation}
d\hat{s}^2=(g_{00}+g_{rr}\dot{r}^2+g_Sh_{ij}\dot{\varphi}^i\dot{\varphi}^j)
dt^2+gds^2_3.
\label{indmet}
\end{equation}
Making use of Eqs. (\ref{radmtn}) and (\ref{angmtn}), one can further 
simplify the $(t,t)$-component of the induced metric.  By defining the 
cosmic time $\eta$ through
\begin{equation}
d\eta^2=-(g_{00}+g_{rr}\dot{r}^2+g_Sh_{ij}\dot{\varphi}^i\dot{\varphi}^j)dt^2,
\label{costime}
\end{equation}
one can put the induced metric (\ref{indmet}) into the following form similar 
to the standard metric for the expanding universe:
\begin{equation}
d\hat{s}^2=-d\eta^2+g(r(\eta))ds^2_3.
\label{expuniv}
\end{equation}
Note, the three-dimensional metric $ds^2_3$ for the case of a source D-brane 
with the non-zero NS $B$-field is not in general that of three-dimensional 
flat Euclidean space, and therefore the metric (\ref{expuniv}) does not 
correspond to the Robertson-Walker metric for a {\it flat} expanding universe 
unlike the previous cases in Refs. \cite{kk,pp1,kim1,kim2,pp2} but rather 
describes the expansion of anisotropic universe.  However, when the probe 
D3-brane is very close and very far away from the source brane, the metric 
$ds^2_3$ approaches the flat Euclidean metric, so one can apply the result 
of the standard flat expanding universe.

To derive an analogue of the four-dimensional Friedman equations for the 
expanding four-dimensional universe on the probe D3-brane, we define the 
scale factor $a$ as $a^2=g$ with the Hubble parameter defined as 
$H=\dot{a}/a$, where the overdot from now on stands for derivative with 
respect to the cosmic time $\eta$.  Since the scale factor is given by 
$a=H^{-1/4}_p=(1+{Q_p\over r^{7-p}})^{-1/4}$ for the source brane solutions 
under consideration in this paper, the four-dimensional universe on the 
probe brane expands [contracts] when the probe brane falls towards [moves 
away from] the source brane. 
The equation describing the evolution of the four-dimensional universe 
on the probe brane is then
\begin{equation}
\left({{\dot{a}}\over a}\right)^2={1\over 4}\dot{r}^2\left({{g^{\prime}}
\over g}\right)^2,
\label{frdeq}
\end{equation}
where the prime stands for derivative with respect to $r$.  The explicit 
expression for $\dot{r}^2$ can be found, case by case, by making use of Eqs. 
(\ref{radmtn}) and (\ref{costime}).  Since the three-dimensional metric 
$ds^2_3$ in Eq. (\ref{expuniv}) is not flat Euclidean for the case under 
consideration in this paper, the RHS of Eq. (\ref{frdeq}) in general is not 
proportional to the effective matter density $\rho_{\rm eff}$ on the probe 
brane.  However, when the probe D3-brane is very close to the source D-brane 
(with the non-zero NS $B$ field), i.e. the IR regime of the dual gauge theory 
on the boundary, and is very far away from the source brane, for which the 
metric $ds^2_3$ approximates to the flat three-dimensional Euclidean space 
metric, one can set the RHS of Eq. (\ref{frdeq}) to be $\approx 
{{8\pi}\over 3}\rho_{\rm eff}$.  We also have
\begin{equation}
{{\ddot{a}}\over a}=\left(1+{g\over{g^{\prime}}}{\partial\over{\partial r}}
\right){1\over 4}\dot{r}^2\left({{g^{\prime}}\over g}\right)^2=
\left[1+{1\over 2}a{\partial\over{\partial a}}\right]
{1\over 4}\dot{r}^2\left({{g^{\prime}}\over g}\right)^2=
-{{4\pi}\over 3}(\rho_{\rm eff}+3p_{\rm eff}),
\label{frdeq2}
\end{equation}
where $p_{\rm eff}$ is the effective pressure.

\section{Brane Cosmology in the Background of D-Brane with the Magnetic 
NS $B$ Field}

In this section, we study the mirage cosmology on the probe D3-brane 
moving in the bulk of the source D$p$-brane with nonzero magnetic NS 
$B$ field.  The four-dimensional universe on the probe D3-brane has 
the space/space noncommutativity.  

The supergravity solution for the source $p$-brane with the rank 2 
magnetic NS $B$ field is given by \cite{bmm,cp,hi,ms,aos}
\begin{eqnarray}
ds^2_{10}&=&H^{-{1\over 2}}_p\left[-fdt^2+h(dx^2_1+dx^2_2)+dx^2_3+\cdots+
dx^2_p\right]+H^{1\over 2}_p\left(f^{-1}dr^2+r^2d\Omega^2_{8-p}\right),
\cr
e^{2\phi}&=&H^{{3-p}\over 2}_ph,\ \ \ \ \ \ \ 
B_{12}=H^{-1}_ph\tan\theta,
\cr
A^p_{01...p}&=&(H^{-1}_p-1)h\cos\theta,\ \ \ \ \ \ 
A^{p-2}_{03...p}=(H^{-1}_p-1)\sin\theta,
\cr
H_p&=&1+{{L^{7-p}_p}\over{r^{7-p}}},\ \ \ \ \ \ 
f=1-\left({{r_0}\over r}\right)^{7-p},\ \ \ \ \ \ 
h^{-1}=\cos^2\theta+H^{-1}_p\sin^2\theta,
\label{magbpbran}
\end{eqnarray}
where $L^{7-p}_p=r^{7-p}_0\sinh^2\alpha_p$.

Substituting the solution (\ref{magbpbran}) into the probe action 
(\ref{probact}), one obtains the Lagrangian density of the form 
(\ref{lagden}) with the coefficients given by
\begin{eqnarray}
A(r)&=&g(g^2h^2+b^2)|g_{00}|e^{-2\phi},
\cr
B(r)&=&g(g^2h^2+b^2)g_{rr}e^{-2\phi},
\cr
D(r)&=&g(g^2h^2+b^2)g_Se^{-2\phi},
\label{magcoefs}
\end{eqnarray}
where 
\begin{equation}
g_{00}=-fH^{-{1\over 2}}_p,\ \  
g=H^{-{1\over 2}}_p,\ \ 
g_{rr}=f^{-1}H^{1\over 2}_p,\ \ 
g_S=r^2H^{1\over 2}_p,\ \ 
b=H^{-1}_ph\tan\theta.
\label{magdef}
\end{equation}  
In the $p=3$ case, we have the following additional term coming from 
the last two terms of the probe action (\ref{probact}):
\begin{equation}
C(r)={1\over{\cos\theta}}(1-H^{-1}_p).
\label{magwzcoeff}
\end{equation}
So, the equations (\ref{radmtn}) and (\ref{angmtn}) for the probe geodesic 
motion take the following forms:
\begin{eqnarray}
\left({{dr}\over{dt}}\right)^2&=&{{|g_{00}|}\over{g_{rr}}}\left[1-
{{|g_{00}|}\over{g_S}}{{g(g^2h^2+b^2)g_Se^{-2\phi}+\ell^2}\over{(C+E)^2}}
\right],
\cr
h_{ij}{{d\varphi^i}\over{dt}}{{d\varphi^j}\over{dt}}&=&
{{g^2_{00}}\over{g^2_S}}{{\ell^2}\over{(C+E)^2}}.
\label{mageqmtn}
\end{eqnarray}

Making use of the equations of the motion (\ref{mageqmtn}), one can simplify 
the induced metric (\ref{indmet}) on the probe D3-brane as follows:
\begin{equation}
d\hat{s}^2=-{{g^2_{00}g(g^2h^2+b^2)e^{-2\phi}}\over{(C+E)^2}}dt^2+gds^2_3,
\label{magindmet}
\end{equation}
where $ds^2_3=hdx^2_1+hdx^2_2+dx^2_3$.  Then, the cosmic time $\eta$, with 
which induced metric (\ref{magindmet}) takes the standard form 
(\ref{expuniv}) of an expanding universe, is defined as
\begin{equation}
d\eta={{|g_{00}|g^{1\over 2}(g^2h^2+b^2)^{1\over 2}e^{-\phi}}\over
{|C+E|}}dt.
\label{magcostime}
\end{equation}
Therefore, the four-dimensional Friedman equation (\ref{frdeq}) 
takes the following form:
\begin{equation}
\left({{\dot{a}}\over a}\right)^2=
{{(C+E)^2g_Se^{2\phi}-|g_{00}|[g_Sg(g^2h^2+b^2)+\ell^2e^{2\phi}]}\over
{4|g_{00}|g_{rr}g_Sg(g^2h^2+b^2)}}\left({{g^{\prime}}\over g}\right)^2.
\label{magfrdeq}
\end{equation}

We now express the RHS of the Friedman equation (\ref{magfrdeq}) in terms 
of the scale factor $a$.  This is achieved by substituting Eqs. 
(\ref{magdef}) and (\ref{magwzcoeff}) into Eq. (\ref{magfrdeq}) and making 
use of the relation $r^{7-p}=L^{7-p}_pa^4/(1-a^4)$.  For the probe D3-brane 
at an arbitrary distance from the source brane, the Friedman equation 
(\ref{magfrdeq}) takes the form:
\begin{eqnarray}
\left({{\dot{a}}\over a}\right)^2&=&\textstyle{{(p-7)^2}\over{16}}
L^{-2}_pa^6\left(\textstyle{{1-a^4}\over{a^4}}\right)^{{2(8-p)}\over{7-p}}
\left[(E\cos\theta)^2a^{2(p-5)}\right.
\cr
& &\left.-\textstyle{{(L^{7-p}_p+r^{7-p}_0)a^4-r^{7-p}_0}\over{L^{7-p}_p}}
\left\{1+\textstyle{{(\ell\,\cos\theta)^2}\over{L^2_p}}
\left(\textstyle{{1-a^4}\over{a^4}}\right)^{2\over{7-p}}a^{2(p-5)}
\right\}\right],
\label{magdensp}
\end{eqnarray}
for $p\neq 3$, and
\begin{eqnarray}
\left({{\dot{a}}\over a}\right)^2&=&L^{-2}_3a^6\left(
\textstyle{{1-a^4}\over{a^4}}\right)^{5\over 2}
\left[\left(E\cos\theta+1-a^4\right)^2a^{-4}\right.
\cr
& &-\left.\textstyle{{(L^4_3+r^4_0)a^4-r^4_0}\over{L^4_3}}\left\{1
+\textstyle{(\ell\,\cos\theta)^2\over L^2_3}\left(\textstyle{{1-a^4}\over{a^4}}
\right)^{1\over 2}a^{-4}\right\}\right],
\label{magdens3}
\end{eqnarray}
for $p=3$.  The constraint (\ref{cnstrnt}) that the RHS of the 
Friedman equations (\ref{magdensp}) and (\ref{magdens3}) are non-negative 
limits the size of the scale factor to be $0\leq a\leq 1$.  
In particular, when the probe D3-brane is in the near horizon region of the 
source brane, for which $H_p\approx L^{7-p}_p/r^{7-p}$, the radial coordinate 
is related to the scale factor as $r^{7-p}\approx L^4_pa^4$.  So, the 
Friedman equation (\ref{magfrdeq}) becomes 
\begin{eqnarray}
\left({{\dot{a}}\over a}\right)^2&\approx&\textstyle{{(p-7)^2}\over{16}}
L^2_pa^{-2{{p-11}\over{p-7}}}\left[(E\cos\theta)^2a^{2(p-5)}\right.
\cr
& &-\left.\textstyle{{L^{7-p}_pa^4-r^{7-p}_0}\over{L^{7-p}_p}}
\left\{1+\textstyle{{(\ell\,\cos\theta)^2}\over{L^2_p}}a^{{8\over{p-7}}
+2(p-5)}\right\}\right],
\label{magnrdensp}
\end{eqnarray}
for $p\neq 3$, and
\begin{eqnarray}
\left({{\dot{a}}\over a}\right)^2&\approx&L^2_3a^{-4}
\left[\left(E\cos\theta+1-a^4\right)^2a^{-4}\right. 
\cr
& &-\left.\textstyle{{L^4_3a^4-r^4_0}\over{L^4_3}}\left(1+
\textstyle{{(\ell\,\cos\theta)^2}\over{L^2_3}}a^{-6}\right)\right],
\label{magnrdens3}
\end{eqnarray}
for $p=3$.  If we consider just the near-horizon geometry (for an arbitrary 
$r$) of the source brane, the constraint (\ref{cnstrnt}) does not restrict 
the allowed value of the scale factor $a$, which thereby taking 
$0\leq a<\infty$.  

We see from the above Friedman equations that the cosmological evolution on 
the probe D3-brane does not have any qualitative difference from the case 
without NS $B$-field
\footnote{It is also observed in Refs. \cite{cai1,cai2,cai3} that there is 
no qualitative difference in thermodynamics of a probe D-brane between the 
cases with and without the magnetic NS $B$-field.} 
studied in \cite{kk}.  Namely, the nonzero magnetic NS 
$B$ field has an effect of just weakening the energy $E$ and the total 
angular momentum $\ell$ in the Friedman equations by the factor $\cos\theta$, 
thereby slowing down expansion of the brane universe.  So, just like the case 
without the NS $B$ field, when the probe D3-brane is very close to the source 
brane (i.e. $a\ll 1$), the Friedman equation is dominated by the term 
$\sim a^{-2(p-11)/(p-7)+2(p-5)}$ for $\ell=0$ and by the term 
$\sim a^{-2(p-11)/(p-7)+8/(p-7)+2(p-5)}$ for $\ell\neq 0$ when 
$p\leq 5$, and by the term $\sim a^{-10}$ for $\ell=0$ and by the 
term $\sim a^{-16}$ for $\ell\neq 0$ when $p=6$.  When the NS $B$ field 
is very large, i.e. $\theta\approx\pi/2$, the Friedman equation 
is dominated by term $\sim a^{-2(p-11)/(p-7)}$, as the rest terms are 
suppressed by the factor $\cos^2\theta\approx 0$.  
Note, when the perfect fluids, which the cosmological models assume 
matter and energy in the universe to be, satisfy the equation of state 
of the form
\begin{equation}
p=w\rho,
\label{eqst}
\end{equation}
where $w$ is time-independent constant, the conservation of energy equation 
can be integrated to give
\begin{equation}
\rho\propto a^{-3(1+w)}.
\label{rhoprbta}
\end{equation}
In the region very close to the source brane, the induced metric 
(\ref{expuniv}) approximates to the Robertson-Walker metric for a flat 
expanding universe, for which case the RHS of the Friedman equation is 
approximately ${{8\pi}\over 3}\rho_{\rm eff}$.  So, the $p=3$ case, for 
which $\rho_{\rm eff}\sim a^{-4}$ (or $w=1/3$) for the large NS $B$ 
field or for the finite $B$ field with $\ell=0$, simulates radiation 
dominated universe.  And the $p=3$ case with finite NS $B$ field and 
$\ell\neq 0$ corresponds to $\rho_{\rm eff}=p_{\rm eff}$, which is 
characteristic of a massless scalar. 
Note, the causality restricts $w$ in the equation of state (\ref{eqst}) 
to be $|w|\leq 1$.  So, the $p=6$ case, for which $\rho_{\rm eff}\sim 
a^{-10}$ (or $w=7/3$) with $\ell=0$ and $\rho_{\rm eff}\sim a^{-16}$ 
(or $w=13/3$) with $\ell\neq 0$, couldn't possibly have been attained 
by real matter on the probe brane, thereby characterizing the effect of 
the mirage matter.   For the $a\approx 1$ case, i.e. when the probe D3-brane 
is far away from the source brane, the leading order behavior of the 
Friedman equation in terms of $\epsilon=1-a^4\approx 0$ is given by
\begin{equation}
\dot{\epsilon}\approx {{|p-7|}\over{16L_p}}\sqrt{E^2\cos^2\theta-1}
\epsilon^{{8-p}\over{7-p}}\ \ \ \ \Longrightarrow \ \ \ \ 
\epsilon\approx (4L_p)^{7-p}(E^2\cos^2\theta-1)^{{p-7}\over 2}\eta^{p-7},
\label{magacl1p}
\end{equation}
similarly as in the case without the NS $B$ field.  
So, the $p<7$ case corresponds to the asymptotically flat solution.  
The subleading corrections (with additional positive powers of $\epsilon$) 
to Eq. (\ref{magacl1p}) are suppressed by the factor $\cos^2\theta$.  From 
Eq. (\ref{magacl1p}), one can also see that the maximum allowed value of 
$\theta$ (thereby the maximum allowed value of the NS $B$ field or the 
noncommutativity parameter) is given by $\cos\theta=1/E$ at later stage of 
cosmological evolution (for which $a\approx 1$).

\section{Brane Cosmology in the Background of D-brane with the Electric 
NS $B$ Field}

In this section, we study the mirage cosmology on the probe D3-brane 
moving in the bulk of the source D$p$-brane with nonzero electric NS 
$B$ field.  The four-dimensional universe on the probe D3-brane has 
the space/time noncommutativity.

The supergravity solution for the source $p$-brane with the rank 2 electric 
NS $B$ field is given by \cite{lr,ms}
\begin{eqnarray}
ds^2_{10}&=&H^{-{1\over 2}}_p\left[h^{\prime}(-fdt^2+dx^2_1)+dx^2_2+\cdots+
dx^2_p\right]+H^{1\over 2}_p\left(f^{-1}dr^2+r^2d\Omega^2_{8-p}\right),
\cr
e^{2\phi}&=&H^{{3-p}\over 2}_ph^{\prime},\ \ \ \ \ \ \ 
B_{t1}=H^{-1}_ph^{\prime}\tanh\theta^{\prime},
\cr
A^p_{01...p}&=&(1-H^{-1}_p)h^{\prime}\cosh\theta^{\prime},\ \ \ \ \ \ \ 
A^{p-2}_{2...p}=(1-H^{-1}_p)\sinh\theta^{\prime},
\label{elecpbrn}
\end{eqnarray}
where $Q_p=r^{7-p}_0\sinh^2\alpha_p$ and $h^{\prime\,-1}=\cosh^2\theta^{\prime}
-H^{-1}_p\sinh^2\theta^{\prime}$.  

Substituting the solution (\ref{elecpbrn}) into the probe action 
(\ref{probact}), one obtains the Lagrangian density of the form 
(\ref{lagden}) with the coefficients given by
\begin{eqnarray}
A(r)&=&(g^3h^{\prime}|g_{00}|-b^2g^2)e^{-2\phi},
\cr
B(r)&=&g^3h^{\prime}g_{rr}e^{-2\phi},
\cr
D(r)&=&g^3h^{\prime}g_Se^{-2\phi},
\label{eleccoefs}
\end{eqnarray}
where
\begin{eqnarray}
g_{00}&=&-fh^{\prime}H^{-{1\over 2}}_p,\ \ \ \ \ \ \ \ 
g=H^{-{1\over 2}}_p,\ \ \ \ \ \ \ \ 
g_{rr}=f^{-1}H^{1\over 2}_p,
\cr
g_S&=&r^2H^{1\over 2}_p,\ \ \ \ \ \ 
b=H^{-1}_ph^{\prime}\tanh\theta^{\prime}.
\label{elecdef}
\end{eqnarray} 
In the $p=3$ case, we have the following additional term coming from the 
last two terms in the probe action (\ref{probact}):
\begin{equation}
C(r)={1\over{\cosh\theta^{\prime}}}(1-H^{-1}_p).
\label{elecwzcoeff}
\end{equation}
So, the equations describing the probe geodesic motion take the following 
forms:
\begin{eqnarray}
\left({{dr}\over{dt}}\right)^2&=&{{gh^{\prime}|g_{00}|-b^2}\over
{gh^{\prime}g_{rr}}}\left[1-{{(gh^{\prime}|g_{00}|-b^2)
(g^3h^{\prime}g_Se^{-2\phi}+\ell^2)}\over{gh^{\prime}g_S(C+E)^2}}\right],
\cr
h_{ij}{{d\varphi^i}\over{dt}}{{d\varphi^j}\over{dt}}&=&
{{(gh^{\prime}|g_{00}|-b^2)^2}\over{g^2h^{\prime\,2}g^2_S}}
{{\ell^2}\over{(C+E)^2}}.
\label{eleceqmtn}
\end{eqnarray}

Making use of the equations of motion (\ref{eleceqmtn}), one can bring the 
induced metric (\ref{indmet}) on the probe D3-brane to the following form:
\begin{equation}
d\hat{s}^2=-\left[{{(gh^{\prime}|g_{00}|-b^2)^2ge^{-2\phi}}\over
{h^{\prime}(C+E)^2}}+b^2g^{-1}h^{\prime\,-1}\right]dt^2+gds^2_3,
\label{elecindmet}
\end{equation}
where $ds^2_3=h^{\prime}dx^2_1+dx^2_2+dx^2_3$.  So, the cosmic time $\eta$ 
is defined through
\begin{equation}
d\eta=\sqrt{{{(gh^{\prime}|g_{00}|-b^2)^2ge^{-2\phi}}\over
{h^{\prime}(C+E)^2}}+b^2g^{-1}h^{\prime\,-1}}dt.
\label{eleccostime}
\end{equation}
Therefore, the four-dimensional Friedman equation (\ref{frdeq}) takes the 
following form:
\begin{equation}
\left({{\dot{a}}\over a}\right)^2=
{{(h^{\prime}|g_{00}|g-b^2)\left[h^{\prime}g_Sg(C+E)^2-(h^{\prime}|g_{00}|g
-b^2)(h^{\prime}g_Sg^3e^{-2\phi}+\ell^2)\right]}\over
{4h^{\prime}g_{rr}g_Sg\left[(h^{\prime}|g_{00}|g-b^2)^2g^2e^{-2\phi}+b^2
(C+E)^2\right]}}\left({{g^{\prime}}\over g}\right)^2.
\label{elecfrdeq}
\end{equation}

We now express the RHS of the Friedman equation (\ref{elecfrdeq}) in terms 
of the scale factor $a$.  For this purpose, we substitute Eqs. (\ref{elecdef}) 
and (\ref{elecwzcoeff}) into Eq. (\ref{elecfrdeq}).  
For the probe D3-brane at an arbitrary distance from the source brane, for 
which the radial coordinate is related to the scale factor as $r^{7-p}=
Q_pa^4/(1-a^4)$, the Friedman equation (\ref{elecfrdeq}) takes the form
\begin{eqnarray}
\left({{\dot{a}}\over a}\right)^2&=&\textstyle{{(7-p)^2Q_pa^6
[(Q_p+r^{7-p}_0)a^4-r^{7-p}_0]}\over{16[Q_pa^4/(1-a^4)]^{{2(8-p)}
\over{7-p}}}}\left[\textstyle{{(Q_p+r^{7-p}_0)a^4-r^{7-p}_0}\over{Q_p}}
-a^8\tanh^2\theta^{\prime}\right]
\cr
& &\times\textstyle{{{{E^2}\over{\cosh^2\theta^{\prime}
-a^4\sinh^2\theta^{\prime}}}-\left\{{{(Q_p+r^{7-p}_0)a^4-r^{7-p}_0}\over{Q_p}}
-a^8\tanh^2\theta^{\prime}\right\}
\left\{a^{2(5-p)}+\ell^2({{1-a^4}\over{Q_pa^4}})^{2\over{7-p}}\right\}}
\over{\left\{{{(Q_p+r^{7-p}_0)a^4-r^{7-p}_0}\over{Q_p}}
-a^8\tanh^2\theta^{\prime}\right\}^2
a^{2(5-p)}+E^2{{a^8\tanh^2\theta^{\prime}}
\over{\cosh^2\theta^{\prime}-a^4\sinh^2\theta^{\prime}}}}},
\label{elecdensp}
\end{eqnarray}
for $p\neq 3$, and
\begin{eqnarray}
\left({{\dot{a}}\over a}\right)^2&=&\textstyle{{Q_3a^6
[(Q_3+r^4_0)a^4-r^4_0]}\over{[Q_3a^4/(1-a^4)]^{5\over 2}}}
\left[\textstyle{{(Q_3+r^4_0)a^4-r^4_0}\over{Q_3}}-
a^8\tanh^2\theta^{\prime}\right]
\cr
& &\times\textstyle{{\left(E+{{1-a^4}\over{\cosh\theta^{\prime}}}\right)^2
{1\over{\cosh^2\theta^{\prime}-a^4\sinh^2\theta^{\prime}}}
-\left\{{{(Q_3+r^4_0)a^4-r^4_0}\over{Q_3}}-a^8\tanh^2\theta^{\prime}\right\}
\left\{a^4+\ell^2({{1-a^4}\over{Q_3a^4}})^{1\over 2}\right\}}
\over{\left\{{{(Q_3+r^4_0)a^4-r^4_0}\over{Q_3}}
-a^8\tanh^2\theta^{\prime}\right\}^2a^4
+\left(E+{{1-a^4}\over{\cosh\theta^{\prime}}}\right)^2
{{a^8\tanh^2\theta^{\prime}}\over{\cosh^2\theta^{\prime}
-a^4\sinh^2\theta^{\prime}}}}},
\label{elecdens3}
\end{eqnarray}
for $p=3$.  As in the case of the source brane with the magnetic NS $B$ 
field, the constraint (\ref{cnstrnt}) that the RHS of the Friedman equations 
(\ref{elecdensp}) and (\ref{elecdens3}) are non-negative limits the size of 
the scale factor to be $0\leq a\leq 1$. 
When the probe D3-brane is in the near-horizon region of the source brane, 
in which case $r^{7-p}\approx Q_pa^4$, the Friedman equation becomes
\begin{eqnarray}
\left({{\dot{a}}\over a}\right)^2&\approx&\textstyle{{(7-p)^2a^2
(Q_pa^4-r^{7-p}_0)}\over{16(Q_pa^4)^{{9-p}\over{7-p}}}}\left[
\textstyle{{Q_pa^4-r^{7-p}_0}\over{Q_p}}-a^8\tanh^2\theta^{\prime}\right]
\cr
& &\times\textstyle{{{{E^2}\over{\cosh^2\theta^{\prime}
-a^4\sinh^2\theta^{\prime}}}-\left\{{{Q_pa^4-r^{7-p}_0}\over{Q_p}}-
a^8\tanh^2\theta^{\prime}\right\}
\left\{a^{2(5-p)}+\ell^2(Q_pa^4)^{2\over{p-7}}\right\}}
\over{\left\{{{Q_pa^4-r^{7-p}_0}\over{Q_p}}-a^8\tanh^2\theta^{\prime}
\right\}^2a^{2(5-p)}+E^2{{a^8\tanh^2\theta^{\prime}}
\over{\cosh^2\theta^{\prime}-a^4\sinh^2\theta^{\prime}}}}},
\label{elecnrdensp}
\end{eqnarray}
for $p\neq 3$, and
\begin{eqnarray}
\left({{\dot{a}}\over a}\right)^2&\approx&\textstyle{{a^2(Q_3a^4-r^4_0)}
\over{(Q_3a^4)^{3\over 2}}}\left[
\textstyle{{Q_3a^4-r^4_0}\over{Q_3}}-a^8\tanh^2\theta^{\prime}\right] 
\cr
& &\times\textstyle{{\left(E+{{1-a^4}\over{\cosh\theta^{\prime}}}\right)^2
{1\over{\cosh^2\theta^{\prime}-a^4\sinh^2\theta^{\prime}}}
-\left\{{{Q_3a^4-r^4_0}\over{Q_3}}-a^8\tanh^2\theta^{\prime}
\right\}\left\{a^4+\ell^2(Q_3a^4)^{-{1\over 2}}\right\}}\over
{\left\{{{Q_3a^4-r^4_0}\over{Q_3}}-a^8\tanh^2\theta^{\prime}\right\}^2a^4
+\left(E+{{1-a^4}\over{\cosh\theta^{\prime}}}\right)^2
{{a^8\tanh^2\theta^{\prime}}\over{\cosh^2\theta^{\prime}
-a^4\sinh^2\theta^{\prime}}}}},
\label{elecnrdens3}
\end{eqnarray}
for $p=3$.

Unlike the case of the source D-brane with the magnetic NS $B$ field, 
the electric NS $B$ field modifies the Friedman equations nontrivially.  
When the probe D3-brane is very close to the source brane, the 
$\theta^{\prime}$-dependent terms give rise to infinite series of subleading 
terms to $(\dot{a}/a)^2$.  
[The leading order behavior of the Friedman equations for the $a\ll 1$ case 
is the same as the magnetic NS $B$ field case.]  These subleading terms 
become important as $a$ increases, and then the cosmological evolution 
becomes qualitatively different from the case without the NS $B$ field when 
$a$ is no longer close to zero.  
When the probe D-brane is far away from the source brane (i.e. the 
$a\approx 1$ case), the leading order behavior of the Friedman equation 
is still given by 
\begin{equation}
\dot{\epsilon}\approx {{|p-7|}\over{16}}Q^{1\over{p-7}}_p\sqrt{E^2-1}
\epsilon^{{8-p}\over{7-p}}\ \ \ \ \Longrightarrow \ \ \ \  
\epsilon\approx 4^{7-p}Q_p(E^2-1)^{{p-7}\over 2}
\eta^{p-7},
\label{elecacl1p}
\end{equation}
similarly as in the magnetic NS $B$ field case, but the nonzero 
electric NS $B$ field produces infinite series of subleading corrections
\footnote{This can be seen by rewriting the $\theta^{\prime}$-dependent 
terms in the Friedman equations by using $\cosh^2\theta^{\prime}-a^4\sinh^2
\theta^{\prime}=1+\sinh^2\theta^{\prime}(1-a^4)=1+\epsilon\sinh^2
\theta^{\prime}$.} 
to Eq. (\ref{elecacl1p}), unlike the magnetic NS $B$ field case, for which 
the nonzero NS $B$ field just suppresses the already existing subleading 
terms by the factor of $\cos^2\theta$.  
Unlike the case of magnetic NS $B$ field, such infinite series of subleading 
corrections to Eq. (\ref{elecacl1p}) do not get suppressed as the electric 
$B$ field increases, i.e. as $\theta^{\prime}$ increases.

\end{document}